\documentclass[aps,prl,twocolumn,showpacs,preprintnumbers]{revtex4}
\usepackage{graphics}
\def \beq{\begin{equation}}
\def \eeq{\end{equation}}
\def \beqa{\begin{eqnarray}}
\def \eeqa{\end{eqnarray}}
\def \ppbar{\langle\overline\psi\psi\rangle}

\begin{document}

\title{Critical behaviour in QCD at finite isovector chemical potential}
\author{Sourendu \surname{Gupta}}
\email{sgupta@tifr.res.in}
\affiliation{Department of Theoretical Physics, Tata Institute of Fundamental
         Research,\\ Homi Bhabha Road, Mumbai 400005, India.}

\begin{abstract}
We report an investigation of criticality in QCD at finite isovector
chemical potential, $\mu_3$, and at zero temperature. At the critical
point, $\mu_3^c\approx m_\pi$, we find that an uncharged scalar and
pseudoscalar and a charged pseudoscalar meson become massless within
the resolution of our measurement. The effective long distance theory
therefore breaks $O(4)$ symmetry by charged pion condensation. This
results in a rising quark number susceptibility. The baryon remains
massive, as indicated by a vanishing baryon number susceptibility.
\end{abstract}
\pacs{11.15.Ha, 12.38.Gc}
\preprint{TIFR/TH/02-04, hep-lat/0202005}
\maketitle

In the last few years there has been a resurgence of interest in the phase
diagram of QCD \cite{wilczek}. Perturbation theory and effective chiral
models have been used to uncover a rich phase structure at finite chemical
potential and low temperatures \cite{litt}. This region of non-vanishing
baryon chemical potential remains out of reach of present lattice
computations, due to the Fermion sign problem \cite{hands}. However, the
same techniques, namely effective chiral models and perturbation theory,
have also been used to investigate the phase structure for non-vanishing
isovector chemical potential \cite{son}. This new direction in coupling
space is amenable to lattice computations, which can therefore test
the various methods applied to computations at finite baryon chemical
potential.

It is clear that the Dirac operator, $M$, is non-negative if
$M^\dag=QMQ^{-1}$ for some operator $Q$. For two-flavour QCD with
isovector chemical potential,
\beq
   M=\left(\matrix{D(\mu_3) & 0\cr0 & D(-\mu_3)}\right)
           \qquad{\rm and}\qquad       Q=\gamma_5\otimes\tau_2,
\label{dirac}\eeq
the action is positive definite provided each flavour satisfies
$\gamma_5$-hermiticity ($D^\dag=\gamma_5D\gamma_5$) and the quark
masses are degenerate \cite{isov}. The phase diagram of this model was
investigated by chiral perturbation theory and by perturbative QCD and a
second order phase transition in the $O(4)$ universality class was found
at $\mu_3=m_\pi$ due to the condensation of charged pions \cite{son}. A
first quenched lattice investigation of this model has been reported
recently \cite{sinclair} and a phase transition due to pion condensation
has indeed been found. When $\det M$ is real and positive, quenching
does not produce the pathologies which can otherwise arise.

We show here that it is possible to determine the symmetry of the
problem, and through universality then predict the critical exponents,
by investigating the spin-flavour content of the massless particles in
the theory. Our quenched computations verify that a second order phase
transition occurs at $\mu_3=m_\pi$. At, and above, the critical point
the scalar $\sigma$, the neutral pion and one linear combination of
charged pions becomes effectively massless. This lends support to the
idea that the effective field theory should have an $O(4)$ symmetry
broken down by giving mass to a charged pion, {\sl i.e.\/}, by pion
condensation. If one is so inclined, this can be further tested by
the usual, more expensive, method of finite-size scaling.  However,
as pointed out in \cite{sinclair}, this is non-trivial since higher
dimensional operators in the lattice theory are important at the cutoffs
which are presently accessible. These could easily influence the effective
exponents seen. Scaling studies become then many-fold more difficult due
to the necessity of taking finer lattices before investigating finite
volume effects.

The partition function of two-flavour QCD is
\beq
   Z = \int{\cal D}U{\rm e}^{-S_W}\det M(m_u,\mu_u)\det M(m_d,\mu_d),
\label{partn}\eeq
where $S_W$ is the Wilson gauge action and each determinant corresponds to
one flavour of quarks. At the moment
we are interested in the case when $m_u=m_d=m$. The chemical potentials
for each flavour can be combined into the baryon and isovector chemical
potentials $\mu_0=3(\mu_u+\mu_d)/2$ and $\mu_3=(\mu_u-\mu_d)/2$ respectively.
It is instructive to invert these relations and write
\beq
    \mu_u = \frac{\mu_0}3+\mu_3,\qquad
    \mu_d = \frac{\mu_0}3-\mu_3.
\label{flav}\eeq
We shall work with $\mu_0=0$ and $\mu_3>0$.  Since the identification
of $u$ and $d$ quarks is arbitrary in the absence of electroweak
interactions, the difference between positive and negative $\mu_3$
is purely conventional. A flip in the sign of $\mu_3$, as preferred in
\cite{son}, can be accomplished by just interchanging our definitions
of $u$ and $d$ quark flavours.

Quantum number densities are defined as
\beq
   n_i \equiv \left(\frac{T}{V_3}\right)\frac{\partial\ln Z}{\partial\mu_i},
\label{number}\eeq
for $i=0$ and $3$. These are linear combinations of the quark
number densities. The quantity $n_3$ is an order parameter for pion
condensation. Other order parameters are the quark condensates $\ppbar$
and $\langle\overline\psi\gamma_5\tau_2\psi\rangle$.  At the expected
phase transition $\ppbar$ would depart from its vacuum value and
$\langle\overline\psi\gamma_5\tau_2\psi\rangle$ would become non-zero. The
added advantage in using $n_3$ is that it also flags the value of $\mu_3$
at which saturation sets in, with every site of the lattice being filled
to capacity with quarks, and continuum physics can no longer be extracted.

The quark number susceptibilities \cite{qsdef},
$\chi_{ij}=\partial n_i/\partial\mu_j$ can also be used to probe the
transition.  The flavour off-diagonal susceptibility is given by
\beqa
   \nonumber
   \chi_{ud} &=& \left(\frac{T}{V_3}\right)\biggl[
      \left\langle{\rm tr}M_u'M_u^{-1}\,{\rm tr}M_d'M_d^{-1}
                 \right\rangle\biggr.\\
      &&\qquad
         \biggl.-\left\langle{\rm tr}M_u'M_u^{-1}\right\rangle\,
             \left\langle{\rm tr}M_d'M_d^{-1}\right\rangle\biggr].
\label{chiud}\eeqa
This differs from the usual $\mu_f=0$ expression \cite{pushan} through
the subtraction of the last term, corresponding to non-vanishing
quark densities.  When a charged pion condensate forms, $\chi_{ud}$
must become non-zero and negative. Expressions for the flavour diagonal
susceptibilities can be read off from \cite{pushan} and modified by
appropriate subtraction of the non-vanishing quark number densities.
As long as the baryon remains massive, $\chi_0=\partial n_0/\partial\mu_0$
should remain zero. $\chi_3$ is a combination of $\chi_0$ and $\chi_{ud}$
and hence gives no extra information.

We explicitly use two staggered quark fields--- one for each component
of the Fermion matrix in eq.\ (\ref{dirac}).
We adopt and extend the usual ``cure'' for Fermion doubling, and write the
two-flavour (one up and one down) determinant as
\beq
   \det M = \left(\det M_u M_d\right)^{1/4}.
\label{stag}\eeq
Consequently, each trace in eqs.\ (\ref{number},\ref{chiud}) comes
with a factor of $1/4$ multiplying it.  There are two ways in which
quark masses enter the problem--- the sea quark mass appears in the
determinants in eq.\ (\ref{partn}) and influence the path integral
measure, while the valence quark mass enters into the operators such as
in eq.\ (\ref{chiud}).  In the quenched approximation the determinants
are set to unity, so the sea quarks drop out of the problem but not the
valence quarks.

It might seem that the use of staggered quarks is a needless complication
since the number of mesons of any given spin is much larger than that
expected in two flavour QCD. However, only four of these are Goldstone
pions at $T=\mu=0$. The others are split from them at any finite
lattice spacing. This trick gives us the ability to bypass some of the
problems seen at coarse lattice spacing. On finer lattices the number
of Goldstone modes for staggered quarks is in any case a problem in
defining two staggered flavours and some other formulation of lattice
quarks is probably better suited to the problem. One other advantage
to using two independent staggered quark fields is that it allows us to
give them unequal masses if we so wish.

The partition function of eq.\ (\ref{partn}) can be expressed through
a transfer matrix--- $Z={\rm Tr}T^N(T,\mu_0,\mu_3)$. The transfer matrix
$T$ propagates field information from one slice of the lattice to another,
and $N$ is the number of slices. There may be several inequivalent ways
to slice the lattice, giving rise to possibly inequivalent transfer
matrices, but the partition function is unique.  The eigenvalues of $T$,
$\lambda_0\ge\lambda_1\ge\cdots$, define masses (or screening masses)
through the formula $1/m_i=\log(\lambda_0/\lambda_i)$.  A critical
point implies massless modes for all the transfer matrices. We utilise
the transfer matrix in the time direction and determine the spin-flavour
content of massless mesons which propagate in the time direction.

Since the theory at finite chemical potential has a conjugate quark
\cite{stephanov}, the symmetries of the transfer matrices must take
this into account. In the problem at hand, there are two staggered quark
fields which are conjugate to each other. The symmetry group at $T=\mu=0$
therefore contains an extra $SU(2)$ factor corresponding to arbitrary
rotations between these two fields. This symmetry is broken at finite
$\mu$. Recall that a chemical potential enters the lattice Dirac operator,
$M$, through the time component of the covariant derivative because every
temporal gauge field is multiplied by the fugacity, $\exp(a\mu)$. As
a result, for finite $\mu_3$, the $Z(2)$ group generated by $t\to-t$
does not commute with the new $SU(2)$. The symmetry of the transfer
matrix can be constructed by flipping the roles of the two quarks when
flipping the sign of time. The irreps can be constructed by taking
$Z_2$ invariant combinations such as $\pi^0_\pm=\bar u\gamma_5 u\pm\bar
d\gamma_5 d$ and $\pi^-=\bar u\gamma_5d$ or $\pi^+=\bar d\gamma_5 u$.
When electroweak corrections are neglected, as is the case here, one
can also use the linear combinations $\pi_\pm=\bar u\gamma_5d\pm\bar
d\gamma_5 u$.  The irrep labels of \cite{golterman} can be retained
with a reinterpretation of the old time-reversal quantum number as the
product of time-reversal and the $Z_2$ flip of the up and down quarks
above \cite{note1}.

\begin{figure}[!ht]
   \scalebox{0.6}{\includegraphics{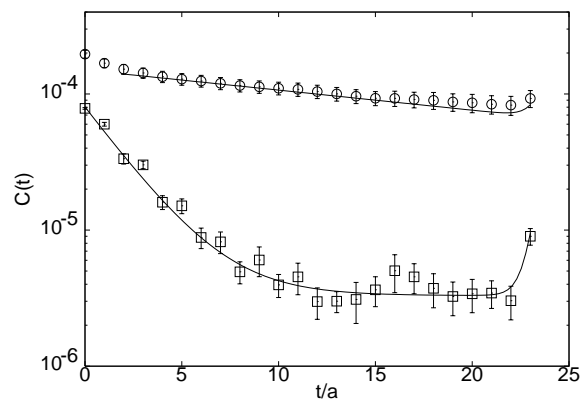}}
   \caption{The correlators for the $\overline dd$ part of $\pi^0$
      (circles) and $\sigma$ (boxes) for $a\mu_3=1$ on a $24\times8^3$
      lattice. The lines are fits to the spin projected part of the
      $\overline uu$ projection and rotated in flavour according to
      the symmetry of the transfer matrix.}
\label{fg.correl}\end{figure}

The importance of identifying the irreps of the transfer matrix follows
from an observation made in \cite{lombardo}: on any finite lattice
an unphysical mode of propagation for a meson is that a quark and an
antiquark travel in opposite directions around the lattice. This finite
size effect is rendered innocuous at $\mu=0$ by proper importance
sampling. However, a chemical potential exacerbates this problem by
encouraging forward propagation of the quark and backward propagation
of the same flavour of the anti-quark.  The correlation function of a
pure flavour state is no longer symmetric in $t$, but has the form
\beq
   C(t)=A{\rm e}^{-mt}+A'{\rm e}^{-m'(N_t-t)},
\label{asym}\eeq
with $m\ne m'$, which seems to indicate the absence of a transfer matrix.
However, by recombining flavours into the irreps of the symmetries of the
transfer matrix we recover the correct symmetric form.

In Figure \ref{fg.correl} we show an example of this. Note that
masses fitted to such pure flavour correlators by the formula in eq.\
(\ref{asym}) are still useful, since the transfer matrix eigenstates
(which are linear combinations of flavours) have the same masses. Tests
of mass fits therefore include the check that the roles of the parameter
pairs ($A,m$) and ($A',m'$) are flipped when fitting the $\bar uu$ and
$\bar dd$ correlators, and that the eigenstates of the transfer matrix do,
in fact, give the lower of these two masses.  Figure \ref{fg.correl} does,
in fact, show one such test, since the correlation functions plotted are
the $\overline dd$ type, and the fits shown are plotted using parameters
obtained from the $\overline uu$ type and flipped.  We have further
checked that local masses also show a plateau close to this mass.

We have simulated quenched QCD on $8^4$ and $12^4$ lattices at
$\beta=5.8941$, which is the critical coupling for the finite
temperature transition in quenched QCD with 6 time slices. The lattice
spacing is then fixed to be $a=1/6T_c$. In the quenched theory, since
$T_c/\Lambda_{\overline{MS}}=1.15\pm0.05$ \cite{scale}, the lattice
cutoff is $a^{-1}=(6.9\pm0.3)\Lambda_{\overline{MS}}$.  The two
lattice sizes differ by 50\% in linear extent. This is necessary to
see and control gross finite volume distortions, but not sufficient to
perform a study with the finesse required to extract critical indices
from finite size effects. We have generated 40 stored configurations
for analysis on the smaller lattice and 70 on the larger. These are
obtained with a Cabbibo-Marinari pseudo-heatbath algorithm where each
$SU(2)$ subgroup imbedded in $SU(3)$ is touched thrice during every
link update.  An initial 5000 sweeps are discarded for thermalisation
and each subsequent stored configuration is separated from the previous
one by 1000 sweeps. The successive stored configurations are therefore
fully decorrelated from each other, rendering the analysis of measurement
errors fairly simple.

Our methods for measuring the number densities and susceptibilities are
the same as in \cite{pushan}. The chiral condensate is measured from the
same matrix inversions. Correlators are obtained with a point source and
projected on to zero 3-momentum by summing over the sink. The spectrum
measurements use standard techniques for staggered quarks, including
fitting to and extracting local masses from spin-projected parts of
the correlators. Since we work on finite lattices, we have a finite
resolution for the identification of massless modes. When a mass reaches
the value $1/aN_t$, the corresponding particle becomes effectively massless
and contributes to the long-distance effective theory.

Since this is a first survey, we chose to work with a reasonably
heavy valence quark mass--- $ma=1/6$. At $T=\mu_3=0$ this gave $m_\pi
a=1.065\pm0.003$ and $m_\rho a=1.323\pm0.007$. The baryon mass was
$m_pa=1.59\pm0.05$. These measurements are compatible with previous
computations at nearby couplings \cite{gottlieb}. Since $m_\pi L\approx12$
on our larger lattice, we expect finite volume distortions to be small
at this quark mass except in the neighbourhood of any critical point,
where, in any case, we expect to see other physics as already explained.

\begin{figure}[!bht]
   \scalebox{0.6}{\includegraphics{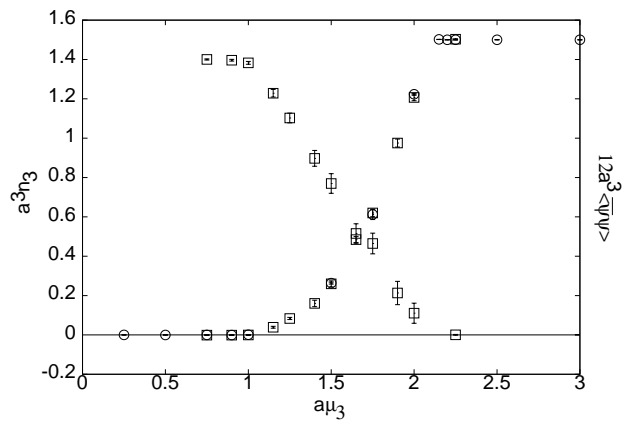}}
   \caption{The order parameters. Boxes denote measurements on $8^4$
      lattices and circles those on $12^4$ lattices.}
\label{fg.order}\end{figure}

Measurements of the order parameters as functions of $\mu_3$ are shown in
Figure \ref{fg.order}. We draw attention to two salient facts. First,
$\ppbar$ and $n_3$ both depart from their vacuum value at the same
$a\mu_3^c\approx1\approx am_\pi$, and, second, the two lattice sizes yield
identical results for the critical $\mu_3^c$. From the flattening of $n_3$
at $a\mu_3\approx2$ it is clear that saturation sets in and continuum
physics can no longer be extracted. The critical point is reached far
before this.  The susceptibility $\chi_{ud}$ is zero at $T=\mu_3=0$
\cite{pushan}. It begins to depart from this value, becoming negative,
soon after $\mu_3^c$. Its behaviour is closely related to
the massless combination of charged pions. Throughout this range of $\mu_3$,
$n_0$ is consistent with zero, indicating that the baryon remains
massive \cite{birse}.

\begin{figure}[!bht]
   \scalebox{0.6}{\includegraphics{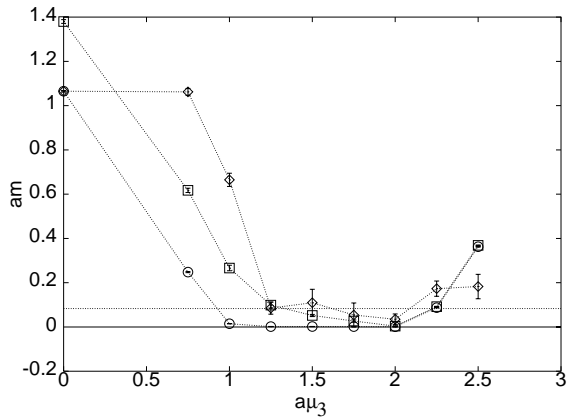}}
   \caption{The spectrum of $\pi^0_+$ (circles), $\sigma_+$ (boxes) and
      $\pi_+$ (diamonds) obtained from temporal correlators on $12^4$
      lattices. The horizontal dotted line indicates the mass resolution
      available on a lattice of this size.}
\label{fg.pion}\end{figure}

Figure \ref{fg.pion} shows the $\mu_3$-dependence of masses of the
$\pi^0_+$ and $\sigma_+$ obtained from zero-momentum correlators. The
masses of the $\sigma_+$ and $\pi^0_+$ drop rapidly from their
$T=0$ values and become nearly massless close to $a\mu_3^c$. They
remain massless till saturation sets in. The $\pi_+$ mass is the
most interesting. It is non-zero at $\mu_3=0$ due to the finite quark
mass and degenerate with the $\pi^0_+$. It does not change appreciably
till $\mu_3^c$, at which point it drops rapidly and becomes effectively
massless soon after. The higher of the two masses in the flavour projected
charged pion sector, corresponds to $\pi_-$. This remains non-zero in
the whole range of $\mu$ studied here. All the critical behaviour seen
in the order parameters is driven by this.  The rise in meson masses in
the unphysical region $a\mu_3>2$ is easy to understand: propagation of
mesons is hard on a saturated lattice. This is also observed in a free
field theory computation at large $\mu_3$.

A few caveats need to be stated. At the resolution available at these
lattice sizes, $a\mu_3^c$ can be observed to lie between 1 and 1.25,
as seen in Figure \ref{fg.order}. Figure \ref{fg.pion} shows that the
mass resolution available on the $12^4$ lattice makes it unnecessary to
improve the bounds on $\mu_3^c$ on this lattice. If one wants to nail
down the critical isovector chemical potential with higher precision,
then one must go to larger lattices, as always. We are presently
exploring the pion-condensed phase in more detail with smaller quark
masses  and larger lattice volumes. These results, and the consequently
more precise determination of $\mu_3^c$ will be presented elsewhere.

There are three notional regions of $\mu_3$. Here we have explored the
vicinity of $\mu_3\approx m_\pi$ where we found that a chiral effective
theory can be written. The region $\mu_3\gg m_p$ is supposed to be
the domain of validity of perturbation theory. This can be reached at
either smaller lattice spacing or quark mass. The region $\mu_3\approx
m_\rho$ is not accessible to either effective theories or perturbative
expansions, and a lattice computation is the only possible tool to
explore this region.  The extraction of the phase diagram in these
qualitatively different regions of $\mu_3$ is a major task, which we
leave to the future.

I would like to  thank D.\ T.\ Son, M.\ Stephanov and R.\ Gavai for
discussions.


\begin{thebibliography}{99}
\bibitem{note1}
   While this work does not use screening correlators, for future work with
   $T\ne0$ and $\mu_3\ne0$ it is interesting to note that the group theory
   of \protect\cite{d4h} carries over by an exact reproduction of this
   argument. This argument also holds for Wilson quarks.
\end{thebibliography}
\end{document}